\documentstyle[11pt,newpasp,twoside]{article}
\input psfig.sty
\def\slHI{H{\,\small\sl I}}
\def\HI{H{\,\small I}}
\def\kms{km s$^{-1}$}

\def\edcomment#1{\iffalse\marginpar{\raggedright\sl#1\/}\else\relax\fi}
\reversemarginpar

\begin{document}
\title{Viewing the circumnuclear medium "through" the radio absorption}
\author{Raffaella Morganti}
\affil{Netherlands Foundation for Research in Astronomy, Postbus 2, 7990 AA
Dwingeloo, The Netherlands \\
morganti@astron.nl}

\begin{abstract}

Observations of radio absorption (free-free and 21 cm neutral hydrogen
absorption) can provide important constraints on the interstellar medium
(either ionised or neutral) surrounding  AGN.  This gas is relevant in the
obscuration of the central regions and, therefore, in producing the
orientation-dependent aspects of the emission from the AGN itself, one of the
key elements of the unified schemes. 
From these observations we can learn: how strong is the evidence for circumnuclear
tori/disks and how often, instead, is the interaction between the radio plasma and
large-scale ISM playing a role; are the tori/disks (when observed) thick or
thin and how important is this gas in affecting the characteristics of radio
sources, especially in their early phase.  Here, I will summarise the recent
results obtained from free-free and \HI\ absorption observations of Seyfert
and radio galaxies, what they can tell us about these issues, and which
questions remain open.

\end{abstract}

\section{Why Radio Absorption}

Obscuration is one of the key ingredients for unified schemes of AGNs.  The
study of the nature of the foreground gas associated with AGN is therefore an
important way to investigate the presence of obscuration and relate it to the
predictions of the ``standard'' unified schemes.  The study of this gas can be
done at radio wavelengths in a number of ways and in this review I will
concentrate on free-free absorption at GHz frequencies and on the 21-cm,
hyperfine transition of atomic hydrogen, seen in absorption against a
radio continuum source.

What can we learn from these observations that is  relevant for unified schemes? The
most interesting features to look for are the
nuclear tori/disks.  Following the original studies (e.g.  Krolik \& Begelman
1988), it has been often assumed that the nuclear tori are composed of
dusty molecular clouds.  It is now clear that, under certain conditions, these
tori/disks can be, at least partly, formed by atomic hydrogen (Maloney et al. 
1996). The torus will be molecular if its pressure exceeds a
critical value that depends on the central source luminosity, the
distance from the source and the attenuating column density between the
central source and the point of interest in the torus.  If the pressure is
below this critical value, the gas is warm ($T\sim10^4$ K) and atomic (Maloney
1998).  Atomic tori/disks can, therefore, be detected in absorption, against a
strong radio continuum source (e.g. the core of the radio source), at the
wavelength of the redshifted \HI. 

On the other hand, the inner edge of the torus is expected to be ionised by
the intense radiation field from the central source.  At the distance of the
inner edge, likely between 0.3 pc and 1 pc from the AGN, a region of depth
$\sim 0.1$ pc is fully ionised at a temperature of $\sim 10^4$ K and density
few $\times 10^4$ cm$^{-3}$.  Such an ionised gas will radiate 
thermal emission and will cause free-free absorption of nuclear radio
components viewed through the torus.  This absorption is responsible for
producing, at radio frequencies, a convex spectrum  with a low-frequency
cut-off. 

In summary, both techniques can tell us about the presence of circumnuclear
tori/disks and their characteristics (e.g.  thickness). 
However, from the \HI\ absorption we can learn more than that about the
nuclear regions of AGNs.  For example, whether gas outside the nuclear
disk-structure can also be relevant for the obscuration of the AGN or whether
kinematically disturbed gas is present, either in the form of {\sl outflow}
due to interaction of the radio plasma with the interstellar medium (ISM) or
{\sl infall}, due to  gas possibly connected with the fuelling of the AGN.

\section {Seyfert galaxies}

\subsection {Large \HI\ disks }

The most complete study of \HI\ absorption in Seyfert galaxies has been done
by Gallimore et al.  (1999).  They observed 13 galaxies and detected
absorption in 9.  Their
main result is that no \HI\ absorption is detected against the central,
pc-scale, radio sources, even in known hidden Seyfert 1's like Mrk 3 or
NGC~1068.  In general, the absorption is detected against extended radio jet
structures but appears to avoid central compact radio sources.  The \HI\
absorption appears to trace gas in rotating disks on the 100 pc-scale, aligned
with the outer disks of the host galaxy, rather than gas associated with the
very central (pc) regions of the AGNs.  Indeed, there is a weak correlation
between the probability to find \HI\ absorption and the  inclination
of the host galaxy.  Among the possible explanations for the result of
Gallimore et al., is that free-free absorption suppresses the background
source so that \HI\ absorption cannot be detected. 

A nice example of a galaxy showing a relatively large-scale disk is Mrk~231,
studied in detail by Carilli et al.  (1998) using the VLBA.  There, a rotating
disk with an east-west velocity gradient of about 110 \kms\ is observed.  Also
in this case the absorption is not toward the radio core on the pc
scale but against a more diffuse radio continuum component seen on the hundred
pc scale.  The only galaxy that does not seems to follow this pattern is
NGC~4151 that appears to be the only case of small scale torus, i.e.  gas on
the pc scale (Mundell et al.  1995).  The column densities of the \HI\
absorption inferred by Gallimore et al.  (1999) are not correlated with those
estimates from the X-ray absorption, but they are systematically lower (the
mean column density from \HI\ absorption is $\sim 10^{21}$ cm$^{-2}$ assuming
a $T_{\rm spin}=100$ K while it is $\sim 10^{22.5}$ cm$^{-2}$ for the hydrogen
column derived from soft X-ray).  This is possibly indicating that the radio
and the X-ray sources are not co-spatial.

\begin{figure}[!h]
\centerline{\psfig{figure=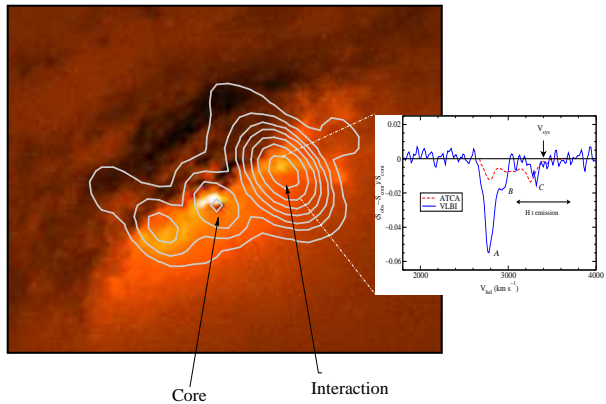,width=8cm,angle=-90}}
\caption{HST image of the central region of the Seyfert galaxy IC~5063
superimposed with the contours of the radio continuum (from the ATCA)
showing  the core and the two lobes. The spectra of the \HI\ absorption 
(obtained with the VLBI and ATCA) against the strongest radio lobe show a
blueshifted absorption, evidence of interaction of the radio plasma with
the interstellar medium (Oosterloo et al. 2000)}
\end{figure}

Gallimore et al.  (1999) found no evidence for major infall/outflow of neutral
hydrogen.  They concluded that in the Seyfert galaxies that they have studied,
the atomic gas is mainly rotating.  However, a clear case of broad \HI\
absorption interpreted as outflow has been found in the Seyfert 2 galaxy
IC~5063 (Morganti et al.  1998, Oosterloo et al.  2000).  In this galaxy, 
highly blueshifted absorption ($\sim 700$ \kms, see Fig.  1) has been
observed.  The absorption is against one of the radio lobes at about 2 kpc
from the nucleus.  In the same position, the kinematics of the ionised gas is
also very disturbed and strong H$_2$ emission has been detected with NICMOS
(Kulkarni et al.  1998) and interpreted as evidence for fast shocks.  IC~5063
is one of the few Seyfert galaxies with a relatively strong radio continuum. 
Thus, similar absorption in other objects {\sl even if present} could be
difficult to detect because of the weaker radio continuum emission.

\subsection{Evidence for free-free absorption}

It has been clearly shown in the case of NGC~1068 (Gallimore et al.  1996, Roy
et al.  1998) that the nuclear continuum source has been attenuated by
free-free absorption. Indeed, thermal emission from the ionised edge of the
accretion disk has been detected in this galaxy (Gallimore et al. 
1997).  Only one other candidate for detection of thermal emission
(NGC~4388) has been found so far (Mundell et al.  2000).  Possible
signatures of free-free absorption due to the ionised inner part of the torus
have been found in a number of Seyfert galaxies.  Pedlar et al.  (1998) report
free-free absorption on scales $< 50$ pc in NGC~4151, while other cases of
free-free absorption in Seyfert down to the scale of $\sim 1$ pc have been
proposed by Wilson et al.  (1998), Roy et al.  (1998, 1999) and Ulvestad et
al.  1999a.  Two examples are shown in Fig.~2. 

It is important to keep in mind, however, that distinguishing between free-free
absorption and synchrotron self-absorption is not always straightforward.  The
spectral shape that they produce is quite similar.  Unlike free-free
absorption, synchrotron self-absorption requires high brightness temperatures
($\sim 10^{9}$ K) due to the presence of a very compact source. 
Unfortunately, the brightness temperatures derived for the observed Seyfert
galaxies are, in many cases, only a lower limit because the sources are still
unresolved at the resolution of the available observations.  Thus, with few
exceptions, not all the data available so far are good enough to completely
rule out synchrotron self-absorption as the cause of the observed absorbed
spectra.

\begin{figure}[!h]
\centerline{\psfig{figure=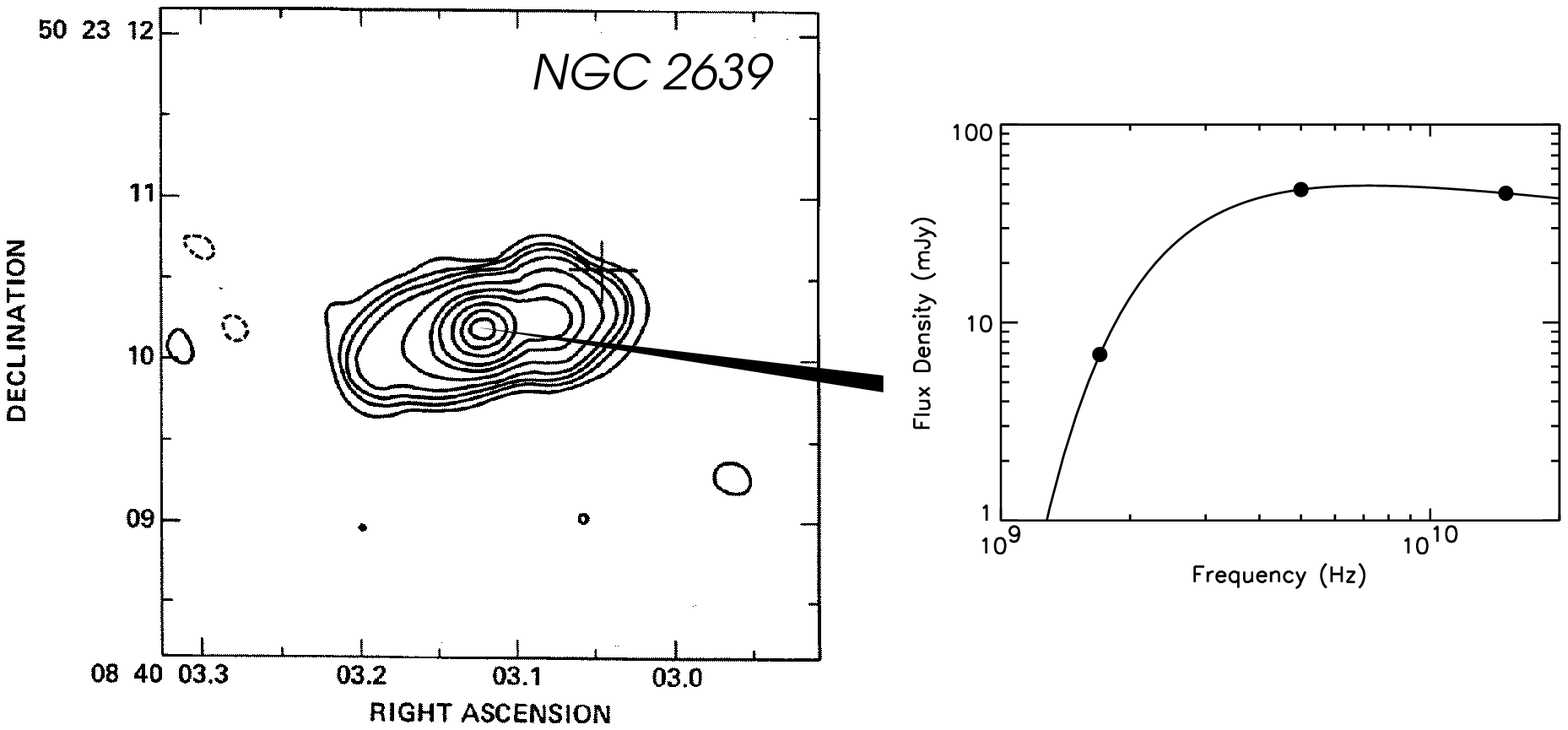,width=8cm,angle=0}
\psfig{figure=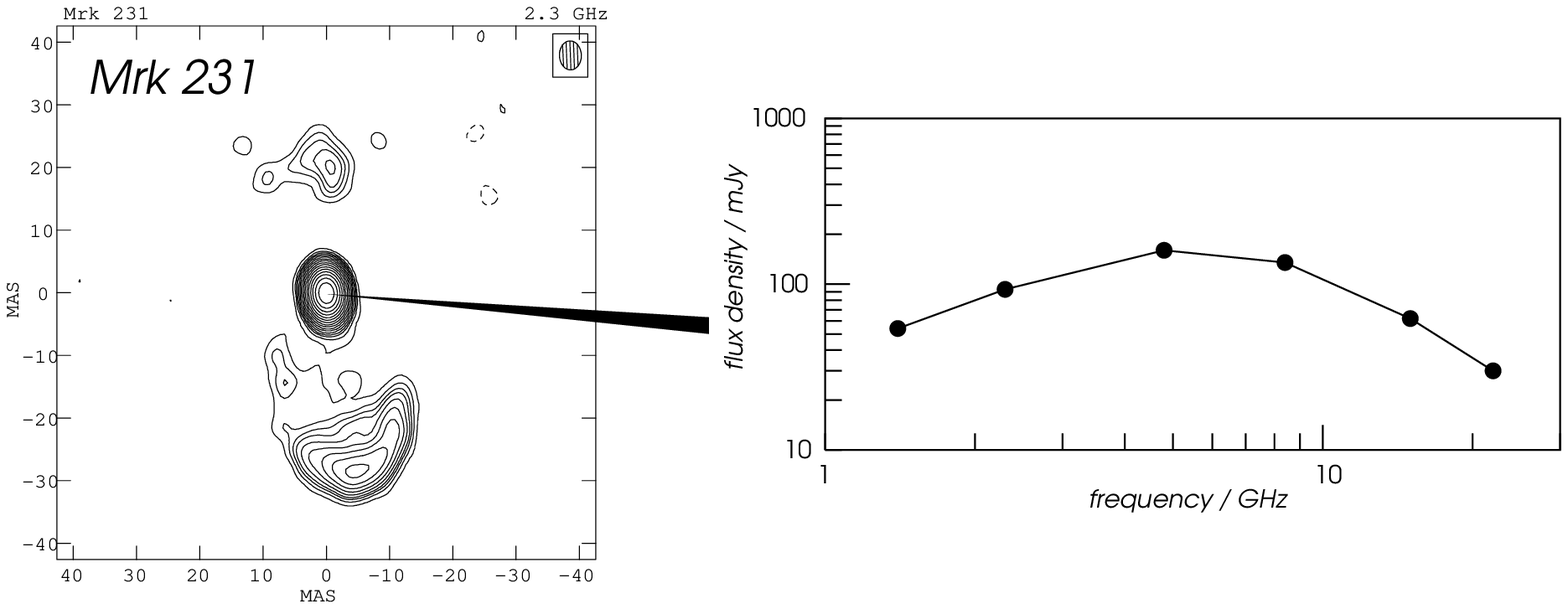,width=9cm,angle=0}}
\caption{Two examples of absorbed spectrum in the Seyfert galaxy. {\sl Left},
NGC~2639, VLA image from Ulvestad \& Wilson and 
spectrum from Wilson et al. 1998 (left);  
Mrk~231 image at 13 cm and spectrum from Ulvestad et al. 1999b (right).
Courtesy of Alan Roy.}
\end{figure}

Supporting the presence of free-free absorption in these objects is the
absence of a counter-jet as observed in a number of Seyfert galaxies that
nevertheless show symmetric radio emission on the larger scale.  In these
cases, Doppler boosting is unlikely to be the cause since the detected
velocity of the jet is low (sub-relativistic, Roy et al.  1999).  Another
indirect argument in support of the free-free absorption interpretation, is
that for a typical column density derived from photoelectric absorption in the
X-ray, $\sim 10^{22.5 - 23}$ cm$^{-2}$, the corresponding densities ($\sim
10^5$ cm$^{-3}$) are of the order of that required to produce free-free
absorption (see e.g.  Wilson et al.  1998).

\section{Looking for nuclear tori/disks in radio galaxies}

\subsection{Tori/Disks in powerful radio galaxies}

For powerful radio galaxies, geometrically thick tori are predicted by unified
schemes.  The combination of obscuration produced by tori/disks and beaming
would explain the lack of broad optical emission lines in some of these
galaxies.  The main problem for the study of \HI\ absorption in this group of
sources is that their radio cores are, on average,  not very strong. 
Because of this, there are only a few cases of extended powerful radio galaxies where
\HI\ absorption has been detected and a detailed study has been done.  

Although  extensive statistics are not available, in a recent study of a flux
limited sample of radio galaxies, Morganti et al. (2001a) found that 
for FR type-II powerful radio galaxies, no \HI\ absorption has been
detected in the few broad line galaxies observed, while three out of four
narrow-line galaxies have been detected (the one non-detection having quite a
high upper limit). To first
order this is consistent with what expected according to the unified schemes
assuming that the \HI\ absorption is due to an obscuring torus. 
However, all the detected sources are compact or small radio galaxies.  This
result confirms what is found  from previous studies (Conway 1997, van
Gorkom et al.  1989) that powerful compact radio sources (i.e.  powerful radio
sources with a Fanaroff-Riley type-II morphology but on the scale of at most
few kpc) are more likely to be detected in \HI\ absorption.  These kind of
radio galaxies are likely to be good screens for detecting the absorption
(Conway 1997), although it is not clear yet whether the high detection rate in
these objects could  also be due to the fact that they are believed to be
young radio sources, and therefore likely to be embedded in a different
(richer?) ISM.

\begin{figure}[!h]
\centerline{\psfig{figure=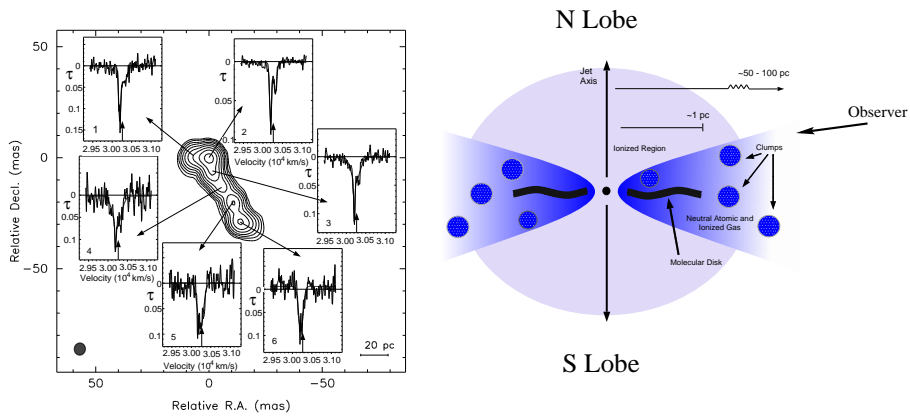,width=12cm,angle=-90}}
\caption{\HI\ absorption profiles toward the radio source 1946+708 overlayed
on the 1.3~GHz continuum contours Peck \& Taylor 2001 (left).  Line of
sight to the jet components (right) in PKS~1946+708 (from Peck et al. 1999). Courtesy of
Alison Peck and Greg Taylor. }
\end{figure}

There are a few cases of well studied radio galaxies that show in more detail
evidence of the presence of tori/disks being the cause of the absorption.
In the extended radio galaxy Cygnus~A,  broad absorption ($\sim 200 -300$
\kms) has been found against the nucleus by Conway \& Blanco (1995).  The
large linewidth of the \HI\ absorption is  quite a common result in the
detected radio galaxies, and it argues against the absorption arising at a large
distance from the nucleus (i.e. kpc-scale).  If it is  gas at large
radii, the expect absorption would be much narrower centred on the systemic
velocity.  From VLBA observation of Cygnus~A, the \HI\ gas seems to be
distributed perpendicular to the radio axis and a velocity gradient is seen
across this distribution.   This suggests a 50 pc scale, rotating,
flattened \HI\ structure  (Conway 1999). 
4C~31.04 is a compact object where a sharp edge in opacity is observed in the
western lobe while in the eastern lobe the opacity is quite uniform (Conway
1997).  This has been taken as a signature of an almost edge-on torus/disk with
thickness $\sim 100$ pc.  
HST observations of 4C~31.04 (Perlman et al. 2001) have confirmed the presence
of a dust disk of the expected dimension as derived from the \HI\ absorption.
The compact radio galaxy PKS~1946+708 (Peck et al.  1999, Peck \& Taylor 2001)
shows \HI\ absorption visible toward the entire $\sim 100$ pc of the continuum
source.  Against the core the absorption is broader ($\sim 300$ \kms) but with
lower optical depth than against the rest of the source (see Fig.~3 and Peck
\& Taylor 2001 for details).  According to Peck \& Taylor (2001), this is in
agreement with the thick torus scenario in which gas closer to the central
engine rotates faster and is much hotter, thus lowering the optical depth. 
The narrow line is explained as coming from gas further out in the torus,
possibly related to an extended region of higher gas density on the order of
at least 80 - 100 pc in diameter. 
The column densities derived for the galaxies described above are $\sim
10^{23}$ cm$^{-2}$ if a T$_{\rm spin}\sim 8000$ K is assumed.  This value is
what expected for purely atomic gas heated by hard X-ray emission. This is
likely to be the condition close to the AGN.  In this case, the column
densities are consistent with what derived from the X-ray observations (one
possible exception being 3C~445, Morganti et al. 2001).

Although most of the detected \HI\ absorption in radio galaxies has been
interpreted as evidence for nuclear tori/disks, there is a growing number of
objects that cannot be explained in this way.  As in the case of the Seyfert
galaxy IC~5063 (see \S 2.1), there are radio galaxies where the \HI\
absorption is associated with gas outflow due to jet-cloud interaction, see
for example 3C~236 (Conway \& Schilizzi 2001) and possibly PKS~1814-63
(Morganti et al. 2000). 
These cases will be discussed in \S 4.2.  A caveat to the interpretation of
the \HI\ absorption, is the accuracy of the systemic velocity derived from the
optical spectra. This is used as the reference velocity 
in the interpretation of the kinematics of the neutral gas.  This will be
discussed later in \S 4.1.

\begin{figure}[!h]
\vskip 1cm
\centerline{Fig.4 here}
\vskip 1cm
\caption{HI absorption detected with the WSRT and the VLA against
the core of two radio  galaxies: NGC 315 and B2 1322+36 (Morganti et al. in prep)}
\end{figure}

\subsection{Thin disks in low power radio galaxies?}

For low power radio galaxies (i.e.  Fanaroff-Riley type I, FRI) the presence of a
thick torus is not so clear yet.  Using HST images, Chiaberge et al.  (1999)
found that unresolved optical cores are commonly present in these radio
galaxies.  The optical flux of these cores appears to be correlated with the
radio core flux, arguing for a common non-thermal origin (synchrotron emission
from the relativistic jet).  All this suggests that the {\sl standard pc-scale
  geometrically thick torus is not present in these low-luminosity radio
  galaxies} (see also Capetti et al. these proceedings) and the cores are,
therefore, mostly unobscured.

In a study of \HI\ absorption of a complete, flux limited, sample of FRI radio
galaxies, Morganti et al.\ (2001a) found a low rate of detection.  Only one of
the 10 FRI galaxies observed was detected in \HI\ absorption.  To first order,
this result is consistent with the idea that the cores of these radio galaxies
are relatively unobscured. 

Recently, \HI\ absorption has been studied in another sample of radio
galaxies. For this sample,  information from HST images about the presence of optical
cores and nuclear dusty disks/lanes is available (Capetti et al.\ 2000, Capetti et al.  these proceedings).  Thus,
the HI observations {\sl aim to correlate the presence (or absence) of \HI\ 
  absorption with the optical characteristics}.  Absorption was detected in
the two galaxies in the sample (see Fig.~4) that have dust disks/lanes and
{\sl no} optical cores (B2 1322+36 and B2 1350+31 (3C293)).  In these cases
the column density of the absorption is quite high ($ > 10^{21}$ cm$^{-2}$ for
T$_{\rm spin} = 100$ K) and the derived optical extinction $A_B$ (between 1
and 2 magnitudes) is such that it can, indeed, produce the obscuration of the
optical cores.  It is worth noticing that B2~1322+36 has a one sided jet on
the VLBI scale.  This is at variance with the idea (see Conway 2001) that
\HI\ absorption is mainly detected 
among the twin-jet sources that have also a counter jet.  Two
galaxies with optical cores were also detected (NGC~315 and
B2~1346+26).  In these cases, however, the derived column density is much
lower ($\sim 10^{20}$ cm$^{-2}$ for T$_{\rm spin} = 100 $ K) and the derived
extinction is of the order of only a fraction of a magnitude. 
If the absence of large obscuration is confirmed for the low-power radio
galaxies, the lack or weakness of broad optical lines compared to other AGN
will have to be explained by something other than obscuration effects.  So
far, broad lines have been {\sl tentatively } found in only a very few cases of
low-power radio galaxies.

\begin{figure}[!h]
\vskip 1cm
\centerline{Fig.5 here}
\vskip 1cm
%\centerline{\psfig{figure=morganti.fig6.ps,width=8cm,angle=-90}}
\caption{The background represent the HST image of NGC~4261. The inset are the
VLBI images (from van Langevelde et al. 2000) of the nucleus at 21 cm (right) and the \HI\ absorption spectrum
observed slightly offset from the nucleus (left). Courtesy of Huib Jan van Langevelde.}
\end{figure}
 
Detailed studies of \HI\ absorption using VLBI data have been done for few
low-power radio galaxies.  The best examples are NGC~4261 (van Langevelde et
al.  2000) and Hydra~A (Taylor 1996).  In both cases, the evidence for a
nuclear disk is based on the fact that the \HI\ absorption is broad ($\sim 80$
\kms).  In Hydra~A, the torus/disk could be quite edge-on (vertical extent
$\sim 30$ pc), consistent with the highly symmetric jets detected on the VLBI
scale.  In NGC~4261 the \HI\ absorption is detected against the counter-jet
(see Fig.~5, van Langevelde et al.  2000) at only 6~pc from the nucleus.

\subsection {Free-free absorption in radio galaxies}

Free-free absorption has been reported for a number of radio galaxies.  A list
of references for the galaxies studied so far can be found in Jones et al. 
2000. 
Two of the best studied objects are NGC~1275 (Walker et al.  2000) and
NGC~4261 (Jones et al.  2000) for which multiple frequencies, high resolution
images were obtained.  In both cases, the free-free absorption is explained as
due to an accretion disk.  In the case of NGC~4261, an electron density (model
dependent) in the inner 0.1 pc of the disk between $3\times 10^3$ and 10$^8$
cm$^{-3}$ is obtained while for NGC~1275 a value of at least $\sim
2 \times 10^4$ cm$^{-3}$ if a thick medium (1~pc) is considered (therefore
rising to higher values for a thinner medium). 

A low frequency spectral turnover has been detected against the core and the
inner part of the receding jet in PKS~1946+708 (Peck et al.  1999).  This can
be due to free-free absorption although, as in the case of Seyfert galaxies,
synchrotron self-absorption cannot be completely ruled out.  If the absorption
corresponds to free-free, it could be part of an obscuring torus, likely the
inner region of the same structure detected in \HI\ absorption.  A similar
situation has been found in Hydra~A where there is evidence (Taylor 1996) that
the core and inner jets are free-free absorbed by ionised gas. According to
Marr et al. 2001,  strong evidence for free-free absorption
are observed against the steep-spectrum lobes of the compact 
radio galaxy 0108+388, ruling out synchrotron self-absorption in this object.

Conway (2001) pointed out that all  radio galaxies where  free-free
absorption has been detected show a parsec scale counter-jet as well as a jet. 
Moreover, the free-free absorption is concentrated against the counter-jet
feature.  This  an indirect evidence that the detected free-free absorption,
if confirmed, is likely to be the result of obscuration from an ionised disk
oriented roughly perpendicular to the radio jet. 

\section{Alternative explanations for the \HI\ absorption}

\subsection{Accurate redshifts \& the \HI\ in two star-bursting radio galaxies}

An important issue for understanding the nature of the \HI\ absorption is how
the systemic velocity of the host galaxy compares with the velocity of the
\HI.  It was already pointed out by Mirabel (1989) how the systemic velocities
derived from optical emission lines can be both uncertain and biased by
motions of the emitting gas.  So far we have found two cases in which this
problem clearly appears: the two compact radio galaxies PKS 1549-79 and
4C~12.50.  In both cases, {\sl two redshift systems} were found from the
optical emission lines, thus making it difficult to define the ``true''
systemic velocity. In PKS 1549-79 and 4C~12.50 the higher ionisation lines
(e.g.\ [OIII]5007\AA) have a significant lower redshift (few hundred \kms\ 
difference) than the low ionisation lines (e.g.\ [OII]3727\AA).  {\sl The
  velocity derived from the low ionisation lines is consistent with the one
  derived from the \slHI}.  This is shown in Fig.~6 for 4C~12.50
where the velocities from the optical lines (Grandi 1977) are compared with
the \HI\ absorption recently obtained from the WSRT (this galaxy was already
known to have \HI\ absorption from the work of Mirabel 1989).

\begin{figure}[!h]
\centerline{\psfig{figure=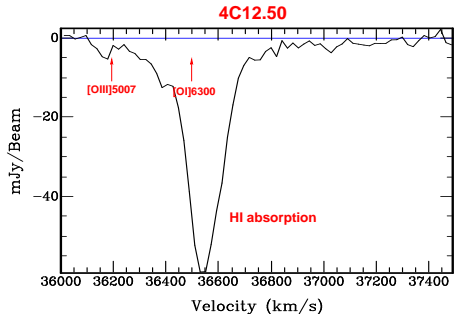,width=6cm}}
\caption{\HI\ absorption of 4C~12.50 observed with the WSRT. The velocity
derived from the optical emission lines ([OIII]500\AA\ and [OII]3727\AA\ from
Grandi 1977)
are marked.}
\end{figure}

In the case of PKS~1549-79, the geometry derived from the radio morphology and
the characteristics of the optical emission lines are not consistent with the
prediction from unified schemes (see Tadhunter et al.  2001 for details).  In
this galaxy the high-ionisation lines are likely to form in a region close to
the central AGN, which is undergoing outflow because of interactions with the
radio jet, while the low-ionisation lines and the \HI\ absorption come from an
obscuring region at a larger distance from the nucleus and thus not so
disturbed kinematically.  PKS~1549--79, as well as 4C~12.50, are likely to be
{\sl young} sources where the nucleus is surrounded by a cocoon of material
left over from the events which triggered the nuclear activity.  Indeed, they
are luminous far-IR sources (4C~12.50 is even classified as ultra-luminous and
it is also rich in CO, Evans et al. 1999) and they have a strong spectral
component from a young stellar population.

%   As the radio source evolves, any obscuring material along the jet
%axis is likely to be swept aside by, e.g., jet-cloud interactions.  Before
%this stage is reached a substantial amount of obscuring material may be
%present along the radio axis.  in which the jets and quasar winds have not yet
%swept aside the warm ISM in the ionisation cones. 

The possibility that the rich medium around a radio galaxy could be
responsible for both the \HI\ absorption as well as being related to the young
stellar population component is supported by the fact that the fraction of
radio galaxies with young stellar population that are detected in \HI\ seems to be
particularly large (so far {\sl all} have been detected) suggesting a link
between the two effects (Morganti et al.  2001a,b).  The starburst component
and the cocoon of absorbing material around the radio lobes can also affect
the way we see (and classify) the AGN, especially at optical wavelengths.
This may be the case  in PKS~1549-79.  This has been recently pointed out also
by the work of Levenson et al.  (2001) for example in the case of the Starburst/Seyfert-1
galaxy NGC~6221. 

{\sl A major implication of these results is that the simplest version of the
unified schemes may not always hold for young, compact radio sources}.

\subsection{Infall and outflow}

In previous studies, the \HI\ absorption was mainly found either at the
systemic velocity or redshifted compared to it (van Gorkom et al.  1990).  As
a result of a number of recent studies, both redshifted and blueshifted (with
respect to the systemic velocity) cases of \HI\ absorption have been found. 
However, it is clear from the above discussion that, in general, accurate
redshifts are needed before we are able to be sure of the interpretation.  At
present, there are only a few cases of clearly redshifted \HI\ absorption. 
One of these cases is NGC~315.  The well known deep and very narrow component
(Dressel et al.  1983) 500 \kms\ redshifted compared to the systemic velocity,
is clearly visible in Fig.~4.  A similar double \HI\ absorption has been found
in 4C~31.04.  These are probably the only two examples, so far, where we are
detecting an infalling cloud that could be physically associated with the
nuclear region (although the case of  NGC~315 is not yet compelling).  Thus, this would
represent a cloud close to the nucleus, falling into it and "feeding" the AGN. 

In some cases, the \HI\ absorption seems to come from gas situated around the
radio lobes and affected by the interaction with the radio plasma (i.e. 
outflow).  The best example is the Seyfert galaxy IC~5063 (described in \S
2.1).  Among the radio galaxies, evidence of jet-cloud interaction affecting
the \HI\ absorption has been found in 3C~236 (Conway \& Schilizzi 2000). In
this galaxy the atomic absorption is associated only with the eastern lobe of
the central 2~kpc mini-double structure.  An other possible example is the
compact radio galaxy PKS~1814-63.  In this galaxy the HI absorption is
observed against the entire radio emission ($\sim 350$ pc, Morganti et al. 
2000).  Most of the \HI\ absorption (with an optical depth as deep as 30\%) is
blueshifted compared to the systemic velocity of the galaxy (at least if we
rely on the redshift available so far).  Thus, this component could be
associated with extended gas, possibly surrounding the lobes and perhaps
interacting/expanding with them.  The superluminal object 3C~216 could be an
other example (Pihlstr\"om et al.  1999). A few more galaxies where the derived
velocity of the \HI\ absorption is blueshifted compared to the systemic
velocity have been found, although more accurate data are still needed to
confirm them.

\section{Summary}

Observations of \HI\ absorption in both Seyfert and radio galaxies appear to
tell us more about the 100-pc scale nuclear disks than about the pc-scale
tori.  In Seyfert galaxies, the 100-pc scale rotating disks traced by the \HI\ 
absorption are aligned with the outer disks of the host galaxy.  Only one case
of absorption on parsec scales has been found.  The nuclear continuum source
is believed to be attenuated by free-free absorption, and signatures of this
have been found in the spectra of a number of Seyfert galaxies, as well as in
the lack of a counter-jet in objects where a sub-relativistic velocity of the
jet has been measured.  However, synchrotron self-absorption as a possible
alternative explanation for the absorbed spectra cannot, in many cases, be
completely ruled out yet.

Evidence for tori on the 100 pc scale has also  been found
in a handful of radio galaxies  from VLBI studies. Only one object shows the absorption against
the counter-jet at a projected distance of 6~pc.  
The low detection rate of \HI\ absorption in low-luminosity radio galaxies, as
found from statistical studies, 
appears to be consistent with thin nuclear disks being characteristic of this
class of radio galaxy.  In radio galaxies, free-free absorption has been
also observed and all the detected objects show a parsec scale counter-jet (as
well as a jet) and the free-free absorption is concentrated against the
counter-jet feature. Thus, also in these objects the free-free absorption is
likely to be the result
of obscuration by an ionised  disk oriented roughly perpendicular to the
radio jet.

Despite these results, it is fair to say that the evidence is sometimes
circumstantial and more objects need to be studied.  Indeed, among both Seyfert
and radio galaxies, there are numerous examples  where the \HI\
absorption traces gas that is not distributed in a disk but is surrounding
the radio lobes and interacting with them (outflow). In a few cases, this
represents 
clouds  falling into the nucleus.  It is therefore extremely
important to obtain accurate systemic redshifts for a correct interpretation of the
\HI\ absorption. 

Finally, the obscuration by a cocoon of gas surrounding the radio lobes
observed some radio galaxies may alter the observed characteristics of the
radio source, making the classification
according to the unified schemes  more complicated .

\acknowledgments

I would like to thank Alison Peck, Alan Roy, Greg Taylor for useful
discussions. Some of the results presented here would not have been obtained
without the help of my collaborators: Tom Oosterloo, Clive Tadhunter, Karen
Wills and Gustaaf van Moorsel.

\end{document}